# An Efficient QOS Based Multimedia Content Distribution Mechanism in P2P Network

| M. Anandaraj [1], | Dr. P. Ganeshkumar[2] | K. P. Vijayakumar [3] |
|---|---|---|
| [1]Associate Professor, | [2]Professor , | [3]Associate Professor, |
| Dept. of Information Technology, | Dept. of Information Technology, | Dept. of Information Technology, |
| PSNA College of Engg. & | PSNA College of Engg. & | PSNA College of Engg. & |
| Technology, Dindigul, India. | Technology, Dindigul, India. | Technology, Dindigul, India. |

*Abstract— Peer-to-peer network is one in which each node in the network can act as a client or server for the other nodes in the network. It allows shared access to various resources such as files, peripherals, and sensors without the need for a central server. Content distribution in the P2P network from server is done by multicasting. Multicasting is the process of sending the data to the multiple designations. This technology is highly efficient for the large scale multimedia content delivery in P2P network where the end peer have identical set of system components. But in reality, the peers have heterogeneous set of requirements for different service levels as well as different service components. The ability to provide differentiated services to each peer with widely varying requirements is becoming important. We need to provide differentiated Services above the existing shared network infrastructure. The solution proposed to solve the above said problem is to provide individualized service to each peer. It focuses on constructing and maintaining an efficient multiple overlay multicast tree structure in the P2P network. The tree maintenance process is governed by two mechanisms called as dynamic reconfiguration driven by peer and less frequent tree maintenance by network status change observation. In this paper new scalable architecture is constructed and analysed based on the above strategies.*

## 1. INTRODUCTION

The Internet is a datagram network, meaning that anyone can send a packet to a destination without having to re-establish a path. Of course, the boxes along the way must have either pre computed a set of paths, or they must be relatively fast at calculating one as needed, and typically, the former approach is used [6][12]. However, the sending host need not be aware of or participate in the complex route calculation; nor does it need to take part in a complex signalling or call setup protocol. It simply addresses the packet to the right place, and sends it. This procedure may be a more complex procedure if the sending or receiving systems need more than the default performance that a path or network might offer, but it is the default model. Adding multicast to the Internet does not alter the basic model. A sending host can still simply send, but now there is a new form of address, the multicast or host group address. Unlike unicast addresses, hosts can dynamically subscribe to multicast addresses and by so doing cause multicast traffic to be delivered to them. Sending multicast traffic is no different from sending unicast traffic except that the destination address is slightly special. This paper organizes as follows. Section 2 overviews the P2P network and its types, section 3 presents the overview of DDHT, section 4 explains the proposed method. Section 5 concludes this paper.

## 2. PEER-TO-PEER NETWORK

A peer-to-peer (abbreviated to P2P) computer network is one in which each peer in the network can act as a peer as well as server for the other peers in the network. This architecture allows the peers to share access to various resources such as files, peripherals, and sensors without the need for a central server. P2P networks can be set up within the home, a business, or over the public network. Each network type requires all computers in the network to use the same or a compatible program to connect to each other and access files and other resources found on the other computer. P2P networks can be used for sharing content such as audio, video, data, or anything in digital format.

### Architecture and Types of P2P network

Peer-to-peer systems often implement an abstract overlay network, built at Application Layer, on top of the native or physical network. This type of overlays is used for indexing and peer discovery and make the P2P system independent from the physical network. Content is typically exchanged directly over the underlying Internet Protocol (IP) network. A pure P2P network does not have the clients or servers. Each peer in the network is at equal rank peers that simultaneously function as both clients and servers to the other peers on the network. This type of network arrangement differs from the traditional client–server model where communication is controlled by a central server. The P2P overlay network consists of all the participating peers as network peers. There are links between any two peers that know each other. We can classify the P2P networks as structured or unstructured.

In structured P2P networks, peers are organized following specific criteria and algorithms, which lead to overlays with specific topologies and properties. They use distributed hash table (DDHT) [2] based indexing, such as in the Chord system (MIT)[1].

Unstructured P2P networks do not impose any structure on the overlay networks. Peers in these networks connect in an ad-hoc fashion based on a loose set of rules. Three categories can easily be seen:
- In pure peer-to-peer systems the entire network consists solely of equipotent peers.
- In centralized peer-to-peer systems, a central server is used for indexing functions and to bootstrap the entire system.
- Hybrid peer-to-peer systems allow such infrastructure peers to exist, often called super peers

**QoS provisioning in P2P network**

For real-time multimedia services inP2P network such as video streaming, providing acceptable end-to-end quality of service (QoS) is imperative[11]. While it is possible to serve diverse and geographically separated users by creating a special service for each user group, this approach is highly inefficient from the standpoint of computing, storage, and networking resources. Multicasting provides significant bandwidth savings and is particularly crucial for the dissemination of live as well as stored high fidelity multimedia content because of the sheer size of the content, the relatively long duration of the session, and the correspondingly high bandwidth requirements.

Heterogeneous multicast groups contain one or more receivers, which would like to get another service or quality of service as the sender provides or other receiver subsets currently use. A very important characteristic which should be supported by Differentiated Services is that participants requesting a best-effort quality only should also be able to participate in a group communication which otherwise utilizes a better service class [15]. The next better support for heterogeneity provides concurrent use of more than two different service classes within a group. Things tend to get even more complex when not only different service classes are required, but also different values for quality parameters within a certain service class.

**Application level multicast for P2P network**

Application-layer multicast has several attractive features: 1) there is no requirement for multicast support in the layer-3 network; 2) there is no need to allocate a global group identifier, such as an IP multicast address; and 3) since data is sent via unicast, flow control, congestion control, and reliable delivery services that are available for unicast can be exploited. A drawback of application-layer multicast is that, since data is forwarded between end systems, end-to-end latencies can be high. Another drawback is that, if multiple edges of the overlay are mapped to the same network link, multiple copies of the same data may be transmitted over this link, resulting in an inefficient use of bandwidth. Thus, the relative increase of end-to-end latencies and the increase in bandwidth requirements as compared with network-layer multicast, are important performance measures for overlay network topologies for application-layer multicast.

Zhichen Xu, et al [5] proposed an efficient overlay application layer multicast infrastructure for mm real time applications based on a combination of milestone clustering and RTT measurements. Zhenhai Duan,et al [7] in a SON purchases bandwidth with certain QoS guarantees from individual network domains via bilateral service level agreement (SLA) to build a logical end-to-end service delivery infrastructure on top of existing data transport networks[4].

### 3. Proposed Framework

We propose the efficient application layer level multicast for delivering the multimedia contents to the all the peers simultaneously in P2P network. In this approach the global view of the system stored in a dynamic distributed hash table (DDHT). It is scalable, fault-tolerant, and administration-free. It updates the information about the peers in the network dynamically. The global view of the system is generated from milestone clustering. Combining the milestone information with a small number of round-trip time (RTT) [8] measurements to locate physically close-by peers, this approach provides very fast, high quality tree construction and adaptation.

Constructing an efficient multicast tree for rich media distribution is complicated by the heterogeneity of peer needs and available resources. An important challenge is to deliver "personalized" end-to-end service that meets the individual needs while keeping the multicast tree structure as efficient as possible [10]. To handle the above conflicting goals, we take a "peer-centric" approach where each peer explicitly specifies the desired service and the QoS requirements[14]. The tree construction algorithm relies on global information maintained in a DDHT, and follows the three intuitive heuristics.
- Existing service paths should be reused to the extent possible.
- New service paths should be created from existing service paths using appropriate transformations to the extent possible.
- New service components should be placed as near as possible to the peers requiring the service.

**P2P Network model**

Each peer contains a separate Distributed hash table and peers are attached to the server. The server may be connected to LAN or WAN. A new peer requests services from the server for distributing media files using distributed hash tables. Server accepting the connection and distributes the media files.

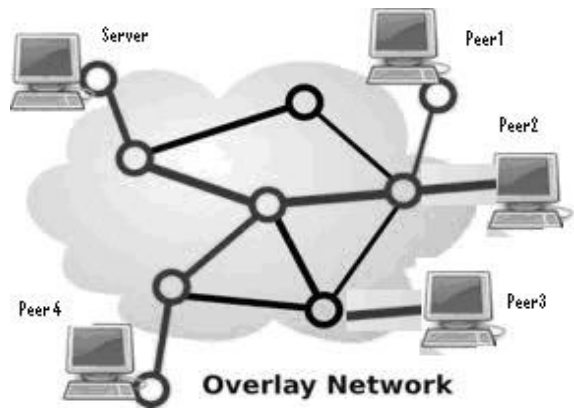

**Figure.1. P2P network with single server**

**Dynamic Distributed Hash Table**

A Dynamic distributed hash table is, as its name suggests, a hash table which is distributed among a set of cooperating computers, which we refer to as peers. Just like a hash table, it contains key/value pairs, which we refer to as items. The main service provided by a DDDHT is the lookup operation, which returns the value associated with any given key. In the typical usage scenario, a client has a key for which it wishes to find the associated value. Thereby, the client provides the key to any one of the peers, which then performs the lookup operation and returns the value associated with the provided key [7]. Similarly, a DDDHT also has operations for managing items, such as inserting and deleting items dynamically. The representation of the key/value pairs can be arbitrary. For example, the key can be a string or an object. Similarly, the value can be a string, a number, or some binary representation of an arbitrary object. The actual representation will depend on the particular application. An important property of DDHTs is that they can efficiently handle large volume of data items. Furthermore, the number of cooperating peers might be very large, ranging from a few peers to many thousands or millions in theory. Because of limited storage/memory capacity and the cost of inserting and updating items, it is infeasible for each peer to locally store every item. Therefore, each peer is responsible for part of the items, which it stores locally.

In the network, every peer should be able to lookup the value associated with any key. Since all items are not stored at every peer, requests are routed whenever a peer receives a request that it is not responsible for. For this purpose, each peer has a routing table that contains pointers to other peers, known as the peer's neighbors. Hence, a query is routed through the neighbors such that it eventually reaches the peer responsible for the provided key.

**P2P Connection establishment**

The tree construction algorithm relies on global state maintained in a DDHT that maps "keys" onto "values." The DDHT is implemented on infrastructure peers that have good availability and network connectivity. The flow shown below describes the process of constructing the multicast tree.

When new peer n wants to join the multicast tree, it computes its own milestone vector and carries out the following steps.
(i) Peer n submits its own milestone vector and its service requirements to the DDDHT infrastructure. The DDDHT infrastructure matches the service path requirements with the stored information to compute a set of candidates close to peer n with which the service requirements can be satisfied directly or a new service path that meets the QoS requirements can be constructed.
(ii) Upon receiving the candidates from the DDHT, the new peer n performs additional measurements by estimating delay and bandwidth between n and the candidate peers. In the case where a new service path needs to be constructed, peer n instructs some candidate peers to perform measurements among themselves to obtain delay and bandwidth information between some of the peers.
(iii) Peer n carries out a series of actions to either reuse an existing service path by attaching to a peer as its child or constructs a new service path.

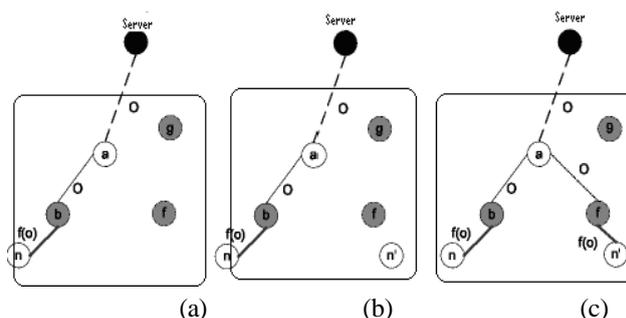

(a)          (b)          (c)
**Figure 2. Connection establishment in P2P network**

Figure 2 exemplifies our tree-construction algorithm. The new peers n and $n^0$ request for service f(O). In Figure 2 (a), the DDHT finds an existing service path (server, ..., a, b) near n and the new peer n attaches to b as its child. In Figure 2 (b), when the new peer $n^0$ wants to join the multicast tree, peers b and n that offer the service are far away from $n^0$. Consequently, peer a that provides the original stream O and peer f that is not on the tree but provides the service, are identified. A new service path (server, ..., a, f, $n^0$) is constructed as shown in Figure 2.

**Milestone based Clustering**

Milestone clustering is based on the intuition that peers close to each other are likely to have similar distances to a few selected milestone peers. Our milestone clustering is based on where a set of well known milestone peers is first identified. The milestone peers measure the RTT among them and use this information to compute a coordinate in a Cartesian space for each of the milestone. To obtain more accurate information, actual RTT measurements are performed against the set of peers that is returned through milestone clustering [9].

DDHT-based overlays, represented by Content Addressable Networks are scalable, fault-tolerant, and administration-free. Their basic functionality is to map "keys" on to "values"[11]. The DDHT stores the global state and is based on CAN, which provides a hash table abstraction over a Cartesian space. The Cartesian space is partitioned into zones, with one or more peers serving as owner(s) of a zone. A key is a point in the space and the owner of the zone that contains the point stores the corresponding value. Since the milestone vectors define a coordinate space, we use the milestone vectors directly as the hash keys.

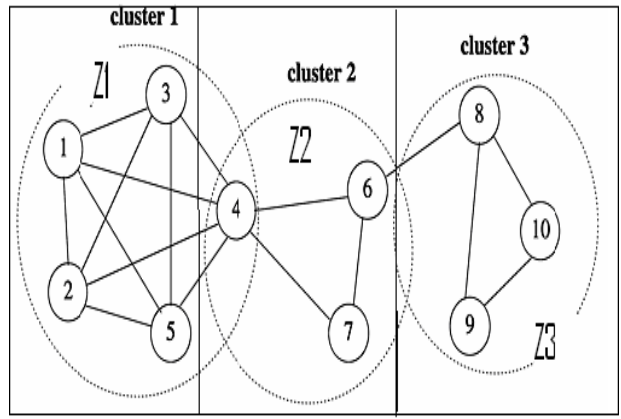

**Figure 3. Storing and retrieving global state on a DDDHT**

Figure 3 illustrates how the global state is stored on a DDHT based on milestone clustering using a two-dimensional CAN[11]. In the figure, the x and y coordinates of the peers are drawn to reflect their milestone vectors. The Cartesian space of CAN is the coordinate space. The coordinate space is partitioned into three zones from Z1 to Z3 with DDHT peers n1 to n4 serving as their owners, respectively [14]. Each DDHT peer owns a CAN [15] zone in which its milestone vector falls into. The information about the tree is stored on the DDHT using their milestone vectors as keys.

The following entries are maintained in the DDHT for each peer in the network

| Entry | Description |
|---|---|
| Peer- ID | Identifier of the node |
| Milestone Vector | It represents the node's position in the physical network |
| Peers -metric | Capacity, load and the services the node provides |
| Path-metric | Characteristics of the path from the root to the node. |

**Table 1. Elements in DDHT**

**Adaptation of Multicast Service Trees**

Driven by the inherent dynamics in the underlying infrastructure, we propose two tree adaptation schemes a just-in-time adaptation to address application quality [13], and a long-term adaptation to address tree quality driven by the network state change observed by the DDHT.

**Dynamic Adaptation of Connection**

This algorithm, based on is driven by application perceived QoS that is impacted not only by fluctuations in the link quality but also by the availability of the requested service. To provide the peers with desired services with reasonable QoS, the connection continuously adapts to the changing conditions and minimizes any service disruption to the peers.

This translates into finding the best location to perform the adaptation and minimizing the latency for each repair. When peer n perceives that the requested service is not available or QoS degrades over its tolerance threshold, it sends a complaint to its parent p in the tree along with its milestone vector and service requirements[16]. If p is not responsive, n switches to a new parent by performing a new join process

**Demand-driven maintenance**

The basic tree construction algorithm is greedy in nature. The order in which the peers join the tree can affect the tree quality.
- When peer n joins the tree, the DDHT maintains information including the requirements of this peer and its upstream peers. Peer n can also specify conditions under which it is interested in getting notified.
- As peers join and leave the system, the DDHT continues to evaluate the predicates of the peers in the system and notifies the appropriate peers when a predicate becomes valid.
- After a peer receives a notification from the DDHT, it makes a local decision as to whether to reconstruct the tree by switching to a new parent.

## 4. PERFORMANCE EVALUATION

We use stress and stretch to measure the quality of the connection. Stretch is defined as the ratio of the sum of the delay attributed to with the tree links to that of a minimal spanning tree. The stress of an overlay multicast tree is the average number of overlay links over a physical link in the underlying topology. Speedup is used to measure the link connection after the connection failure for the proposed algorithm.

**No. of Peers Vs Stretch**

This following graph shows the tree quality measurement. The Y axis represents no of peers in the network and Y axis represents stretch. Our proposed method takes less amount time to make a connection among the peers.

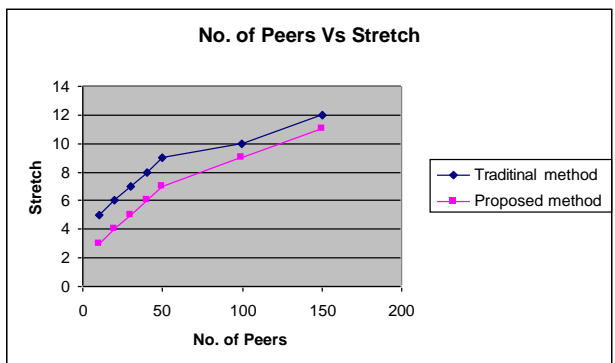

**Figure 4. No. of Peers Vs Stretch**

**No. of Peers Vs Speed up**

This following graph represents the measurement of tree construction time. The no of peers are represented by y axis and the time is represented by the X axis. It shows speed up to reconstruct the peer to peer network after the any link failure.

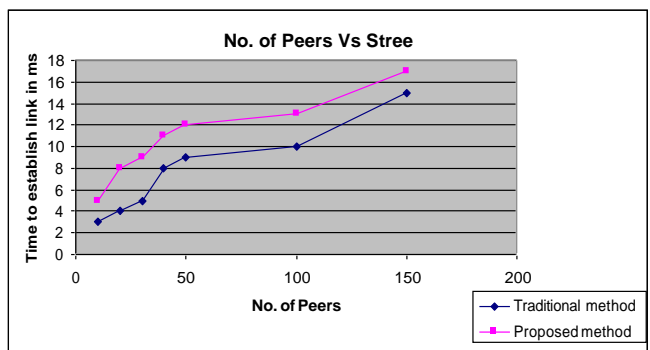

**Figure 5. No. of Peers Vs Speed up**

## 5. Conculsion

We presented new framework which provides composite individualized QoS based services to peers. It balances the need for an efficient infrastructure that maximizes utilization of resource at the same time. The mechanism for delivering content to multiple peers in the network rests on providing a global view of the system stored in a dynamic distributed hash table (DDHT). The global view is generated from milestone clustering. Combining the milestone information with a small number of RTT measurements to locate physically close by peers, the approach provides very fast, high quality tree construction and dynamic tree adaptation. The future work focuses on improving the accuracy of the milestone clustering scheme.

**Authors Biography**

**M.Anandaraj** received his B.E. degree in Computer Science and Engineering from Madurai Kamaraj University (India) in 2003, his M.E. degree in Computer and Communication from Anna University Chennai (India) in 2007, and doing Ph.D. degree in Information and Communication Engineering at Anna University, Chennai. He has been working as an Associate Professor in the Department of Information Technology at PSNA College of Engineering and Technology, India since June 2003. His research interests include computer networks, particularly in network optimization, multicast algorithm design, network game theory and network coding.

**Dr. P. Ganeshkumar** is a Professor in the Department of Information Technology at PSNA College of Engineering and Technology, India since December 2002. He received his B.E degree in Electrical and Electronic Engineering from Madurai Kamaraj University , India in 2001, his M.E. degree in Computer Science and Engineering from the Bharathiyar University (India), and his Ph.D. in Information and Communication Engineering at Anna University, Chennai. His research interests include ADHOC Network, Wireless Networks, Distributed Systems.

**K.P.Vijayakumar** received his B.E. degree in Information Technology from Madurai Kamaraj University (India) in 2003, his M.E. degree in Computer Science and Engineering from Anna University Chennai (India) in 2007, and doing Ph.D. degree in Information and Communication Engineering at Anna University, Chennai. He has been working as an Associate Professor in the Department of Information Technology at PSNA College of Engineering and Technology, India since June 2003. His research interests include computer networks, particularly in network optimization, Wireless network, ADHOC Network, network security                                                  .